\documentclass[]{aa}  
\usepackage{graphicx}
\usepackage{txfonts}
\usepackage{natbib}
\def\micron{$\mu$m}
\def\deg{$^\circ$}

\def\vinfty{$v_{\infty}$}
\def\vsini{$v\sin i$}
\def\Mdot{$\dot{\rm M}$}
\def\kms{${\rm km}\,{\rm s}^{-1}$}
\def\Lsun{L$_{\odot}$}
\def\Rsun{R$_{\odot}$}
\def\Rstar{R$_*$}
\def\Lstar{L$_*$}
\def\Msun{M$_{\odot}$}
\def\Msunyr{\Msun\,yr$^{-1}$}
\def\Ha{H$\alpha$}
\def\Brg{Br$\gamma$}
\def\cmfgen{{\sc cmfgen}}
\begin{document}
\authorrunning{O. Chesneau et al.}
\titlerunning{The variable stellar wind of Rigel }
\title{The variable stellar wind of Rigel probed at high spatial and spectral resolution
\thanks{Based on observations collected at the European Southern Observatory (ESO Programmes 078.D-0355 and 084.D-0393) and at the Observatorio Cerro Armazones (OCA) in Chile.}
}
%
\author{
  O.~Chesneau\inst{1}
  \and
   A.~Kaufer\inst{2}
 \and
  O.~Stahl\inst{3}
  \and
  C.~Colvinter\inst{2}
  \and
  A.~Spang\inst{2}
  \and
  L.~Dessart\inst{4}
  \and
  R.~Prinja\inst{5}
   \and
  R.~Chini\inst{6,7}
}
\offprints{
  O.~Chesneau\\
  \email{Olivier.Chesneau@oca.eu}
}
\institute{
Laboratoire Lagrange, UMR7293, Univ. Nice Sophia-Antipolis, CNRS, Observatoire de la C\^ote d'Azur, F-06300 Nice, France
  \and
  European Southern Observatory, Alonso de Cordova 3107, Casilla 19001, Santiago 763-0355, Chile
  \and
  ZAH, Landessternwarte, K\"onigstuhl 12, D-69117 Heidelberg, Germany
  \and
  Laboratoire d'Astrophysique de Marseille, Universit\'e de Provence, CNRS, 38 rue Fr\'ed\'eric Joliot-Curie, F-13388 Marseille Cedex 13, France
 \and
   Department of Physics \& Astronomy, University College London, Gower Street, London, WC1E 6BT, UK
 \and
	Instituto de Astronom\'{\i}a, Universidad Cat\'{o}lica del Norte,
Avenida Angamos 0610, Antofagasta, Chile
\and
Astronomisches Institut, Ruhr-Universit\"at Bochum,
Universit\"atsstra\ss{}e 150, D-44801 Bochum, Germany
}
%
%
%
\abstract
{Luminous BA-type supergiants can be observed in distant galaxies and are potential accurate distance indicators. The impact of the variability of the stellar winds on the distance determination remains poorly understood. 
}
{Our aim is to probe the inhomogeneous structures in the stellar wind using spectro-interferometric monitoring. 
}
{We present a spatially resolved, high-spectral resolution ($R=12000$) K-band temporal monitoring of $\beta$\,Orionis (Rigel, B8\,Iab) using {{\sc Amber}} at the VLTI. Rigel was observed in the Br$\gamma$\ line and its nearby continuum once per month during 3 months in 2006-2007, and 5 months 2009-2010. These unprecedented observations were complemented by contemporaneous optical high-resolution spectroscopy. We analyse the near-IR spectra and visibilities with the 1D non-LTE radiative-transfer code CMFGEN. The differential and closure phase signal exhibit asymmetries that are interpreted as perturbations of the wind. 
}
{A systematic visibility decrease is observed across the Br$\gamma$\ line indicating that at a radius of about $1.25$\,\Rstar\ the photospheric absorption is filled by emission from the wind. During the 2006-2007 period the Br$\gamma$\ and likely the continuum forming regions were larger than in the 2009-2010 epoch. Using CMFGEN, we infer a mass-loss rate change of about 20\% between the two epochs. We further find time variations in the differential visibilities and phases. The 2006-2007 period is characterized by noticeable variations of the differential visibilities in Doppler position and width and by weak variations in differential and closure phase. The 2009-2010 period is much more quiet with virtually no detectable variations in the dispersed visibilities but a strong S-shape signal is observed in differential phase coinciding with a strong ejection event discernible in the optical spectra. The differential phase signal that is sometimes detected is reminiscent of the signal computed from hydrodynamical models of corotating interaction regions. For some epochs the temporal evolution of the signal suggests the rotation of the circumstellar structures.
}
{}

\keywords{
  Techniques: high angular resolution --
  Techniques: interferometric  --
  Stars: mass-loss --
  Stars: rotation --
  Stars: individual (HD 34085 (Rigel)) --
  Stars: circumstellar matter}
\maketitle
%
%
\section{Introduction}

The discovery of systematic, patterned variability in the stellar winds of luminous hot stars is important for our understanding of wind dynamics and stellar structure. 
Numerous time resolved observations of O stars like the {\sc Feros} campaigns on \object{HD\,152408} (O8\,If) \citep{2001A&A...367..891P} and particularly the {\sc Iue Mega} campaign on \object{HD\,64760} (B0.5\,Ib) \citep{1995ApJ...452L..61P, 1995ApJ...452L..53M} and on \object{$\zeta$\,Puppis} (O4\,I(n)f) \citep{1995ApJ...452L..65H} as well as the extended optical {\sc Heros} campaigns over typically 100 consecutive nights on late B- and early A-supergiants \citep{1996A&A...314..599K, 1996A&A...305..887K,  1997A&A...320..273K} and early B-hypergiants \citep{1997A&A...318..819R} have led to a radically new view of hot-star winds. These campaigns have shown that the winds of hot stars are continuously variable on time scales associated with processes on the stellar surface. Consequently, the steady state, spherically symmetric descriptions usually used to model and interpret stellar wind diagnostics (cf. e.g. \citet{2006A&A...445.1099P}) can only provide some sort of mean representation of these outflows. One of the key questions for understanding the nature and origin of the wind variability is its connection to the stellar photosphere.
For OB-type supergiants like \object{HD\,64760}, \object{HD\,152408}, \object{$\zeta$~Puppis}, and the BA-type supergiants like \object{Rigel} (\object{$\beta$\,Orionis}, B8\,Ia) and \object{Deneb} (\object{$\alpha$\,Cygni}, A2\,Ia), cyclic modulation of the stellar wind plays an important role. Obviously, the winds are modulated by a mechanism related to the photospheric rotation, presumably patches on the stellar surfaces produced either by non-radial pulsation (NRP) patterns \citep{2007A&A...463.1093L} or magnetic surface structures ("spots" or extended "loops") \citep{2008A&A...483..857S}. The spatial structures in the photospheres locally change the lower boundary condition of the stellar wind, which in turn can cause localized structural changes in the wind. These structural changes modulate the observed stellar wind profiles as they are dragged through the line of sight to a distant observer by underlying rotation of the star.  For the prototypical B0.5\,Ib supergiant \object{HD\,64760}, \citet{1996ApJ...462..469C} have modeled such structured winds in the context of corotating interaction regions (CIRs). \citet{2006A&A...447..325K} have presented strong evidence that multiperiodic NRPs are the direct source of the regularly spaced wind structures in this star.
Stellar activity has deep consequences for the interpretation of the spectra of hot massive stars, the macroturbulent broadening being a good example of such an issue  \citep{2010ApJ...720L.174S, 2011IAUS..272..212S, 2009A&A...508..409A, 2002A&A...383.1113D}. Stochastic variability is also present related to small-scale clumping of the radiative wind \citep{2005A&A...437..657D}.
Clearly, to make progress understanding these new phenomena and their impact on the fundamental process of mass-loss via stellar winds -- and therefore on stellar evolution scenarios -- , the mechanism(s) responsible for dividing the stellar wind into structured low and high density regions must be determined.

Recent advances in optical interferometry open up a completely new window for the exploration of the wind structures of hot supergiants. For the first time, {\em direct} measurements of the spatial extension of the circumstellar structures as function of wavelength across a wind-sensitive spectral line like \Brg\ become feasible. 
Currently only two instruments provide spectrally dispersed interferometric observables with a spectral resolution power larger than $R =10000$ and a spatial resolution around a milliarcsecond: the {\sc Vega} visible recombiner at {\sc Chara} \citep{2009A&A...508.1073M} and the {\sc Amber} near-IR recombiner \citep{2007A&A...464....1P} at the {\sc Vlti}. These instruments are well suited to study the wind activity of the brightest BA supergiants in our vicinity in wind-sensitive spectral lines such as \Ha\ or \Brg\ \citep{2010A&A...521A...5C}. 
\object{Deneb} and \object{Rigel} were observed by Chesneau et al.~with {\sc Vega} in 2009. The extension of the \Ha\ 
line forming region of both stars was accurately measured and compared with \cmfgen\ models \citep{1998ApJ...496..407H}. Moreover, clear signs of activity were observed in the differential visibility and phases. These pioneering observations, based on recombination of only two telescopes at a time were limited but show the path for a better understanding of the spatial structure and temporal evolution of localized ejections using optical interferometry.

The bright late-type B supergiant \object{Rigel} (= \object{$\beta$\,Orionis} = \object{HD\,34085}) (B8\,Ia) with a K-band magnitude of 0.2 and a stellar diameter of 2.76\,mas \citep{2008poii.conf...71A} is an ideal target for a more in-depth spectro-interferometric study of the variability of hot supergiants. 
The photosphere and wind variability of \object{Rigel} have been extensively studied by \citet{1996A&A...314..599K, 1996A&A...305..887K,  1997A&A...320..273K} by means of spectroscopic time series in the optical. The red-to-blue variations of the \Ha\ emission and absorption features are reminiscent of the variations of Be-stars and indicate the presence of circumstellar structures preferably localized in the equatorial plane and close to the stellar surface between 1 and 2 stellar radii.  The time scales of the variations are of the order of several weeks to months and compatible with the estimated stellar rotation period of 100\,days considering that several active regions are present at the same time. Extreme events like the high-velocity absorptions (HVA) \citep{1996A&A...314..599K} have indicated that circumstellar structures can be stable for several rotational cycles.
Temporal variations in the near-infrared regime remain to date mostly unexplored. However, the \Brg\ line is expected to play an important role in this respect since the near-IR continuum is formed closer to the wind launching region near the photosphere. Therefore, near-infrared lines such as \Brg\ are expected to provide complementary information to the well-established optical spectral variability.

In this work we report on a more ambitious extensive spectro-interferometric monitoring of \object{Rigel} using the high-spectral resolution mode of {\sc Amber}. The {\sc Amber} observations were collected in two observing campaigns in 2006-2007 and 2009-2010 covering the full observing season of \object{Rigel} over some five months each. The {\sc Amber} observations were complemented by quasi-simultaneous high-spectral resolution optical monitoring. 
Interestingly, the 2009-2010 {\sc Amber} observations performed from the Modified Julian Date (defined as Julian Date - 2400000.5, hereafter MJD) MJD\,55139 until MJD\,55300 partially overlap with the global \object{Rigel} monitoring campaign -- known as the 'Rigel-thon' -- involving long-term spectroscopic monitoring, Microvariability and Oscillations in STars ({\sc Most}) space photometry, and spectropolarimetry \citep{2012ApJ...747..108M, 2011IAUS..272..212S}. Owing to the brightness of the star, stringent constraints on the possible existence of magnetic field were provided up to a limit of the individual measurements as low as 13\,G and exclude large dipolar fields in the range of 20-50\,G. The {\sc Most} data also fostered some theoretical developments on the asteroseismology of this star \citep{2012ApJ...749...74M}.

The paper is structured as follows. In Sect.~\ref{sec:datared} we present the optical and near-infrared spectroscopic and interferometric data and their reduction. Sect.~\ref{sec:continuum} analyses the available interferometric continuum information. Sect.~\ref{sec:timeseries} describes the spectral variability in the \Ha\ and \Brg\ lines using time series of spectra of \object{Rigel} obtained for the first time quasi-simultaneously in the optical and the near-infrared. In Sect.~\ref{sec:visibility} and \ref{sec:phase} the interferometric observables visibility and phase are analysed to describe the size and structure of the \Brg\ line-forming region and the observed possibly rotating circumstellar structures. After a summary of the results and a discussion in Sect.~\ref{sec:discussion} we close with our conclusions in Sect.~\ref{sec:conclusions}.
\section{Observations and data reduction}
\label{sec:datared}
\subsection{Optical spectroscopy}
The optical observations of \object{Rigel} during the 2006-2007 campaign were carried out with the high-resolution echelle spectrograph {\sc Feros} \citep{2000SPIE.4008..459K} at the ESO/MPG 2.2-m telescope at La Silla.  A total of 183 spectra with a resolving power of $R=48000$ and a wavelength coverage from $3600-9200$\,\AA\ were collected in 20 nights between October 3, 2006 and March 31, 2007. 
For the 2009-2010 campaign the {\sc Beso} (Bochum Echelle Spectrograph for OCA) spectrograph at the 1.5-m Hexapod-Telescope at the Observatario Cerro Armazones (OCA), Chile, was used. {\sc Beso} is basically identical to {\sc Feros} and was built by the Ruhr-Universit\"at Bochum and Landessternwarte Heidelberg \citep{2011MNRAS.411.2311F}. A total of 79 spectra were obtained between October 6, 2009 and April 7, 2010.
Table~\ref{tab:log_obsOptical} provides the dates of the individual observations. {\sc Feros} obtained sequences of up to 20 spectra of 1\,s exposure time to achieve a typical $S/N$-ratio of $400$ per combined spectrum (at 5400\,\AA). {\sc Beso} obtained a typical $S/N$-ratio of $250$ per spectrum from single 100\,s exposures.
Flatfield and wavelength-calibration exposures have been obtained with the instrument-internal Halogen and Thorium-Argon lamps at the beginning of the respective nights. All spectra have been reduced semi-automatically with ESO-{\sc Midas} using the dedicated {\sc Feros} context as described e.g.~in \citet{1999ASPC..188..331S}.
All spectra have been reduced to barycentric velocities and have been normalized to the stellar continuum using the very stable instrument response curve and low-order fits to clean stellar continuum points.
Throughout this paper, all velocities are given with respect to the laboratory wavelengths of the corresponding lines of interest. For this purpose the wavelengths of the respective lines have been corrected by a systemic velocity of $v_\mathrm{sys} =+18$\,km\,s$^{-1}$ for \object{Rigel} \citep{1996A&A...305..887K}.

\subsection{Near-IR spectroscopy and interferometry}

\object{Rigel} was observed at the ESO Paranal Observatory with the Astronomical Multi BEam Recombiner (hereafter {\sc Amber}), the near-infrared instrument of the {\sc Vlti} \citep{2007A&A...464....1P}. {\sc Amber} operates in the J, H, and K bands with spectral resolutions of 35, 1500, and 12000, combining either three 8.2-m Unit Telescopes (UTs) or 1.8-m Auxiliary Telescopes (ATs). The observations were carried out on a monthly basis in two campaigns, the first from December 30, 2006 until March 8, 2007, and the second from November 4, 2009 until April 13, 2010. The observations were performed in the high spectral resolution mode ($R=12000$, i.e., 25\,\kms\ per 2-pixel resolution element) and the spectral range from $2.147 - 2.197$\,\micron\ in order to resolve the observable velocity fields of \object{Rigel} in the \Brg\ line at $2.1661193$\,\micron. 
The 2006-2007 campaign was performed with the UTs due to the sensitivity limitations of {\sc Amber} in high-resolution mode at that time. Significant improvements of the {\sc Amber} instrument allowed us to move in 2009 from the UTs to the ATs which then provided data of comparable quality. 
The interferometric observations were carried out by ESO in Service Mode. Due to the significant investment of {\sc Vlti} observing time to obtain an interferometric time series of \object{Rigel} over a complete observing seasons from October to March, the frequency of observations was limited to one measurement (i.e. one 'triplet' of baselines) per month --- despite the known faster variability timescales of \object{Rigel}'s circumstellar environment. Some of the monthly measurements were performed at several days interval, implying that the atmospheric conditions specifically described to perform these observations optimally were not met and that the first dataset of such consecutive observations is of worse quality. 
However, these datasets were still analysed in the same way to test the errors and biases that can be expected from the data reduction.
The log of the {\sc Amber} observations of \object{Rigel} has been compiled in Tab.~\ref{tab:log_obsNIR}.

{\sc Amber} records spectrally dispersed fringes on the detector and therefore provides wavelength-dependent measurements of the size, shape and with sufficient spectral resolution also kinematics of the source corresponding to about 8 independent spectral channels through the line. Three telescopes were systematically recombined, providing a data set consisting in three spectra, three dispersed visibilities, three differential phases, as well as one dispersed closure phase. 
The visibility is the normalized amplitude of the Fourier transform of the intensity distribution of the stellar source in the plane of the sky from which information on the size and shape of the object can be retrieved through some model-fitting analysis. We note that no image inversion is possible with this very limited dataset.

The atmospheric turbulence blurs the fringes at a fast rate so that it is not possible to measure a phase signal directly for each baseline, but the three telescope recombination provides two observables, the differential and closure phase that can be related to the phase of the source. The differential phase represents the wavelength-dependent phase of the source relative to a reference channel. The reference channel is dominated by the continuum signal and the differential phase should consequently be observed with a mean value of zero degree. Strong departures from this mean value may be observed at some spectral channel centred in a line, indicating that the line-forming region at this velocity channel is offset with respect to the continuum. The closure phase is the sum of the phases around a closed triangle of baselines (i.e., $\phi_{12} + \phi_{23} + \phi_{31}$). This quantity is theoretically not affected by the atmospheric turbulence and the value for a spherical source should be zero. A systematic departure at the level of a few degrees from this value is observed due to instrumental defects that affect similarly the science and calibrator sources. After removal of this bias, a detected non-zero closure phase can be interpreted as an evidence of the asymmetrical nature of the source.

Observations of \object{Rigel} were systematically performed together with the observations of a calibrator to allow to correct for the instrumental transfer function and other biases to the visibilities and the closure phases. Finding a good calibrator under these constraints of observation is difficult. \object{Rigel} is a bright, nearby late-B supergiant exhibiting a limited apparent angular diameter given the {\sc Amber} and {\sc Vlti} spatial resolution. A good calibrator would have been a smaller source with a similar flux, that is, an even earlier source, moreover located at the immediate vicinity of \object{Rigel}. The calibrator \object{31\,Ori} (\object{HD\,36167}) is a K5\,III giant and was primarily chosen due to its brightness (K=0.8 versus K=0.2 for \object{Rigel}).
Since the estimated angular diameter of \object{31\,Ori} is 3.56$\pm$0.06\,mas \citep{2002A&A...393..183B} and therefore slightly larger than the one of \object{Rigel}, the accuracy of the absolute visibility calibration remains limited. However, such a measurement is not the main goal of this study.

The data was processed with the the standard {\sc Amber} data reduction software \citep[hereafter {\sc Amber} DRS, \texttt{amdlib} 2.2, see for instance][]{2007A&A...464...29T}.
Some procedures were developed to clean the data that are now inserted by default in the latest version \texttt{amdlib} 3 (better detector cosmetics). The best frames with S/N$\geq$1 were selected and frames affected by high piston excluded. 

\begin{figure}
   \includegraphics[width=0.34\textwidth, angle=-90]{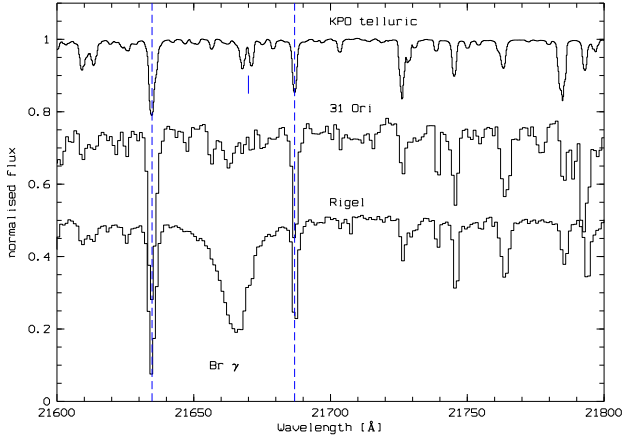}
   \caption{Wavelength calibration of the {\sc Amber} observations. A telluric spectrum from the Kitt Peak Observatory (KPO) is shown in comparison with {\sc Amber} spectra of \object{Rigel} and \object{31\,Ori} around the \Brg\ line. The dashed vertical lines indicate the positions of the two telluric lines used as wavelength reference. The telluric line doublet at 2.1670\,\micron\ seen in the KPO spectrum does not feature in the \object{31\,Ori} spectrum. Therefore, the \Brg\ line profile of \object{Rigel} is {\it bona fide} undisturbed by telluric lines.}
   \label{fig:wave_cal} 
 \end{figure}

In this correction process, realistic error estimates were derived from the data, including the uncertainties on the calibration stars diameters, the instrument and atmosphere transfer function instabilities and the fundamental noises. The study of the transfer function showed that $V^2$ uncertainties were often as large as 10-15\%. Therefore, in the following only differential observables are used in the interferometric analysis.
The data from the UTs are heavily affected by vibrations of the optical surfaces in the optical train that decrease significantly the level of the transfer function. 
In addition, the data recorded during the 2006-2007 campaign are affected by a time and wavelength dependent, high-frequency beating in the spectra, raw visibilities, differential phases, and closure phases. This Fabry-Perot effect has been identified to be caused by an optical component in the {\sc Amber} instrument. A Fourier transform of the observable quantities shows that at first order the beating is periodic. The data were corrected by suppressing the periodic peak in the Fourier domain. This correction removes the majority of the beating effect from the data.
The zeropoint of the {\sc Amber} wavelength scale is not stable and drifts with time. To obtain an accurate wavelength calibration of the {\sc Amber} data, the spectral shifts were measured relative to two telluric water lines at 2.163477 and 2.168687\,\micron\ \citep[ HITRAN]{1992JQSRT..48..469R} that bracket the \Brg\ line. The result of the correction is shown in Fig.~\ref{fig:wave_cal} for \object{Rigel} and \object{31\,Ori} in comparison to a convolved telluric K-band spectrum from Kitt Peak Observatory. The wavelength calibration error of the reduced {\sc Amber} data is estimated to be $\Delta\lambda = 1\cdot10^{-4}$\,\micron\ corresponding to less than $15$\,\kms. The spectral resolving power is measured from unblended telluric lines with $25$\,\kms\ equivalent to $R=12000$.
Eventually, all wavelength scales have been reduced to barycentric velocities and corrected by the systemic velocity $v_\mathrm{sys} =+18$\,km\,s$^{-1}$ of \object{Rigel}.
\section{Continuum information}
\label{sec:continuum}

At the {\sc Hipparcos} distance of $d=237^{+57}_{-38}$\,pc, the radius of \object{Rigel} is $70$\,\Rsun, and at the distance $d=360$\,pc adopted by \citet{2006A&A...445.1099P}\footnote{We note however that they make use of the biased value of Hanbury Brown of 2.55\,mas, leading to an estimate of $99$\,\Rsun.}, this corresponds to $106$\,\Rsun. The width of the \Brg\ line forming region can be estimated to be $15-25$\,\Rsun.

The absolute visibilities of {\sc Amber} are not accurate enough to provide strong constraints on the angular diameter.
Due to the high spectral resolution of the data only a narrow spectral band is covered limiting the range of spatial frequencies for each observation. However, given the comparatively large number of independent visibility measurements in our data set the respective angular diameters have been determined with some statistical significance: 
from the good quality data of the 2009-2010 campaign the angular diameter of \object{Rigel} in the K-band is estimated with a mean of 2.77$\pm$0.08\,mas and a median of 2.79\,mas. The lower quality data of the 2006-2007 campaign provided an angular diameter systematically larger with a 3.01$\pm$0.28\,mas and a median value is 2.94\,mas (this is due to one bad measurement).

The errors are estimated from the standard deviation of the diameter measurements performed at each date of observation but do not take into account the systematical bias introduced by the calibrator \object{31\,Ori}. It is not excluded that a true variability of the apparent diameter of the source may contribute to this scatter, since the \Brg\ line-forming regions seems also larger during 2006-2007 (see following section). The 2009-2010 results are consistent with the {\sc Ionic}  measurement at the {\sc Vlti} of 2.8$\pm$0.1\,mas \citep{2004A&A...424..719L} and with the best estimate to date of 2.76$\pm$0.05\,mas from {\sc Fluor} at {\sc Chara} using K-band measurements with baselines reaching up to 300\,m \citep{2008poii.conf...71A}. 
The 2006-2007 and 2009-2010 continuum closure phases are compatible with a value of zero. No significant departure of the continuum closure phase averaged over the available spectral band is observed.  The 2009-2010 continuum closure phases range from -6.1$\pm$3.9\deg\ to 2.4$\pm$2.9\deg, with a mean value of -0.7\deg and a median of -0.2\deg. The standard deviation of the  closure phase is 1.8\deg, the median value 1.4\deg. The 2006-2007 continuum closure phases have a mean value of 0.9$\pm$2\deg\ and a median of 1.1\deg.
\section{Spectral time series information}
\label{sec:timeseries}
\subsection{\Ha\ and \Brg\ time series}
 \begin{figure}[h!]
   \centering   \includegraphics[width=0.48\textwidth]{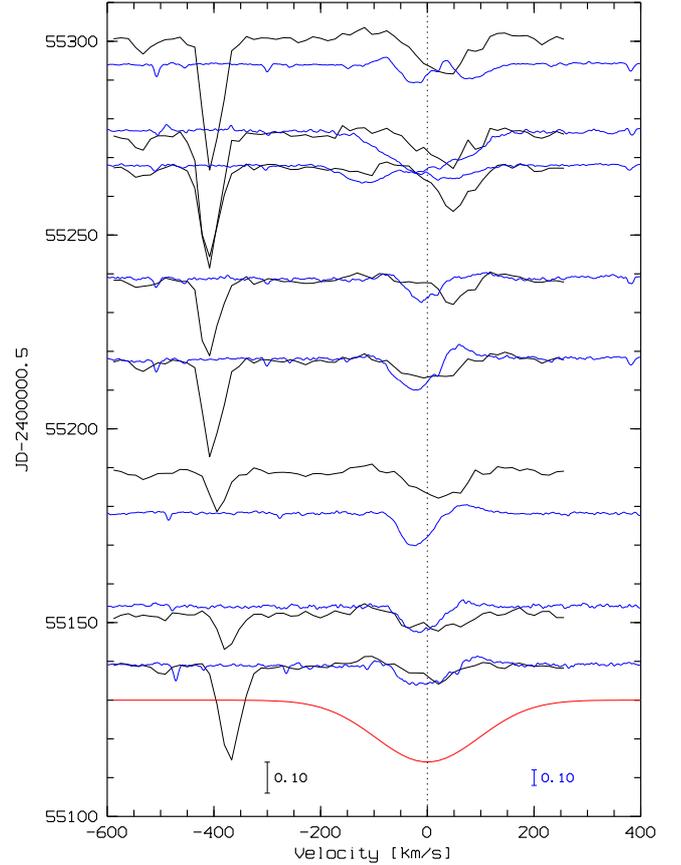}
   \caption{2009-2010 \Brg\ (black) and \Ha\ (blue) time series of \object{Rigel}. The \Brg\ spectra are shown as the difference to an artificially created photospheric spectrum shown at MJD\,55130 (red). For \Ha\ only the spectra taken closest in time to the \Brg\ spectra are shown. The normalised flux scale for \Brg\ is enlarged by a factor of two relative to the one of \Ha\ to match the strength of profile variability of the two lines. The strong telluric water absorption line at $-400$\,\kms\ is not resolved and provides an estimate of the spectral resolution of the {\sc Amber} spectra around \Brg. The complete 2009 time series are shown in Fig.~\ref{fig:Rigel_tsHa2009} and Fig.~\ref{fig:Rigel_tsBrg2009}. }
   \label{fig:Rigel_tsHaBrgdiff2009} 
 \end{figure}

The \Ha$\lambda6562.818$\,\AA\ and \Brg$\lambda21661.19$\,\AA\ recombination lines are the most sensitive lines accessible to optical and near-infrared high-resolution spectrographs and interferometers like {\sc Feros}/{\sc Beso} and {\sc Amber} to probe circumstellar material close to the stellar photosphere of \object{Rigel}. The complex variability of the circumstellar environment has been documented by \citet{1996A&A...305..887K} through extensive monitoring of the \Ha-line variability and was interpreted as rotational modulation of the lower wind regions of the star by complex circumstellar structures. No corresponding data sets existed so far in the near infrared. The {\sc Amber} spectro-interferometric monitoring presented in this work allows for the first time to compare time series of \Ha\ and \Brg\ spectroscopic line profiles of \object{Rigel}.

Figure~\ref{fig:Rigel_tsHaBrgdiff2009} shows a selected subset of the available \Ha\ and \Brg\ time series. All available \Brg\ spectra taken during the 2009-2010 campaign are shown as difference of the recorded \Brg\ spectra and an artificial Gaussian photospheric line profile to allow more direct comparison of the \Brg\ line profiles with \Ha. The photospheric line profile was constructed from a Gaussian profile with a full width at half maximum (FWHM) of 22.8\,\kms\ to match the wings of the \Brg\ profiles and a central depth of 0.8 times the continuum to reproduce the typical central depth of the \Ha\ line profiles. To allow more direct comparison of \Brg\ and \Ha\ only the \Ha\ spectra closest in time to the \Brg\ spectra are shown (the complete \Ha\ and \Brg\ time series from the 2006-2007 and 2009-2010 campaigns are shown in Figs.~\ref{fig:Rigel_tsHa2006}, \ref{fig:Rigel_tsBrg2006}, \ref{fig:Rigel_tsHa2009}, and \ref{fig:Rigel_tsBrg2009} in the appendix and are discussed in the following sections).

Figure~\ref{fig:Rigel_tsHaBrgdiff2009} shows that the {\sc Amber} spectra despite their considerably lower spectral resolution than the {\sc Beso} spectra mostly resolve the line-profile variations displayed in the \Brg\ line profile. It further suggests that the line-profile variability observed in 
\Ha\ and \Brg\ are qualitatively very similar. Both \Ha\ and \Brg\ display red-to-blue variations of emission and absorption features which are reminiscent of the variations of Be-stars and indicate the presence of circumstellar structures in the equatorial regions of the star as established earlier from extended optical time series by \citet{1996A&A...305..887K}. The line-profile variability is primarily caused by variable additional emission superimposed to the underlying photosphere -- wind profiles. However, it is further clear that the \Brg\ line profiles do not necessarily match the corresponding \Ha\ profile: while \Ha\ and \Brg\ often appear very similar with matching emission and absorption components like at MJD\,55217 and MJD\,55267, the profiles appear with opposite emission and absorption components at MJD\,55238. A closer inspection of the full time series in Figs.~\ref{fig:Rigel_tsHa2009} and \ref{fig:Rigel_tsBrg2009} does not reveal any obvious time delay between the evolving line profiles of \Ha\ and \Brg\ as it could be expected from features propagating through the photosphere -- wind transition zone at the base of the wind. 
It is therefore concluded that the \Ha\ and \Brg\ lines are both probing well the photosphere -- wind transition zone at the base of the wind of \object{Rigel} but either at different stellar radii and volumes or through different (local) line-formation processes (see Fig.~\ref{fig:CMFGEN}).
\subsection{2006-2007 campaign}
The \Ha\ time series from the 2006-2007 campaign is shown in Fig.~\ref{fig:Rigel_tsHa2006}. The \Ha\ profiles display most of the time a double-peaked profile with blue and red emission peaks with maxima at $-70$\,\kms\ and $+70$\,\kms, respectively, i.e., at velocities higher than the estimated projected equatorial rotation velocity of $v \sin i = \pm 36$\,\kms.  
The most notable feature of the time series is the blue shifted high-velocity absorption (HVA) at MJD\,54037 with a blue-edge velocity of the absorption feature of $-390$\,\kms\ and a maximum depth of 12\% of the continuum at $-127$\,\kms. HVAs have been observed in \object{Rigel} and other BA supergiants before and were first described in  \citet{1996A&A...305..887K}. The 2006 HVA is only recorded by one single {\sc Feros} spectrum with the previous spectrum at MJD\,54019 and the next at MJD\,54071, i.e., several weeks before and after the event. \Ha\ displays unusually strong inverse P-Cygni profiles at MJD\,54019 before the event and between MJD\,54087 and MJD\,54117 after the event, the latter with a strong inverse P-Cygni profile slowly developing back into the more regular double-peaked profile as described above. Due to the sparse sampling of the 2006-2007 time series no clear links can be established between the sequence of line profiles (inverse P-Cygni profile, HVA, decaying inverse P-Cygni profile, double-peaked profile). The even fewer \Brg\ spectra recorded with {\sc Amber} and shown in Fig.~\ref{fig:Rigel_tsBrg2006} do not allow to fill the time gaps. Unfortunately, no {\sc Amber} spectra were recorded near the HVA event. The \Brg\ spectra between MJD\,54099 and 54108 map the inverse P-Cygni profiles in \Ha\ by showing a red-shifted absorption core of the line profile. Subtracting the artificial photospheric profile as described above reveals inverse P-Cygni profiles in \Brg, too. 
So far, the best-documented HVA events of \object{Rigel} in 1993 and 1994 (Figs.~1 and 2 in \citealt{1996A&A...305..887K}) also indicate the appearance of inverse P-Cygni profiles before and after the observation of HVAs. In the 1993 event the inverse P-Cygni profile was recorded some 19 days before the deepest HVA profile, i.e., with the same time difference as in 2006.
The blue-edge velocity of $-390$\,\kms\ observed in the 2006 HVA is considerably higher than the ones observed in 1993 and 1994 with $-238$\,\kms\ and $-278$\,\kms, respectively, and considerably higher than the terminal wind velocity estimated from UV spectra of the Mg {\sc ii}$\lambda\lambda$2795,2803 lines with a lower limit of $-229$\,\kms. \citet{1991VA.....34..249G} reported on discrete absorption components extending to $-400$\,\kms\ in \object{Rigel}.    
\subsection{2009-2010 campaign}
The \Ha\ and \Brg\ time series from the 2009-2010 campaign are shown in Figs.~\ref{fig:Rigel_tsHa2009} and \ref{fig:Rigel_tsBrg2009}. The \Ha\ profiles display from MJD\,55110 to MJD\,55240 double-peaked profiles with red-to-blue variations of emission and absorption features as most commonly seen in \object{Rigel}. 
At MJD\,55264 an absorption feature with a blue-edge velocity of $-200$\,\kms\ becomes discernible at blue-shifted velocities and rapidly develops  into a broad absorption feature crossing the line profile from blue to red within the next two weeks reaching a red-edge velocity of $+100$\,\kms\ at MJD\,55280.
Interestingly, \Brg\ already shows red-shifted absorption at $+50$\,\kms\ at MJD\,55238 (Fig.~\ref{fig:Rigel_tsBrg2009} and \ref{fig:Rigel_tsHaBrgdiff2009}). The red-shifted absorption even increased in strength (depth) by MJD\,55267 and is still visible at MJD\,55275.
\section{Differential visibilities and the Br$\gamma$ line forming-region}
\label{sec:visibility}
Figures~\ref{fig:Rigel_vis2006}, and Figs.~\ref{fig:Rigel_vis2009}--\ref{fig:Rigel_vis2010} show the measured differential visibilities across the Br$\gamma$ line from the 2006-2007 and 2009-2010 data set, respectively. For each observation three differential visibilities are shown corresponding to the three baselines formed by the ATs. Given the uncertainties of the continuum visibility measurements with {\sc Amber}, the continuum visibilities are set according to a fixed value of $2.76$\,mas for the angular diameter of the photosphere of \object{Rigel} \citep{2008poii.conf...71A}.

The interpretation of the interferometric observables (visibilities and phase) requires the support by theoretical models of the stellar photosphere and wind. The stellar parameters used for the model calculations are taken from \citet{2006A&A...445.1099P} and \citet{2008A&A...487..211M}. The Mg{\sc ii} resonance lines suggest terminal wind speeds of $\sim$230\,\kms\ for \object{Rigel} and the projected rotational velocity is low, at about 36$\pm$9\,\kms\ \citep{1996A&A...305..887K}. All spectra and the dispersed visibilities of the Br$\gamma$ line are computed for a spectral resolution $R=12000$.

The modelling approach is extensively described in \citet{2010A&A...521A...5C}. We used a similar modelling approach but with the model parameters listed in Tab.~\ref{tab:cmfgen}. The mass-loss rate was varied to cover the values 1, 2, 4, 6, 7, 8, 9 and $10 \times 10^{-7}$\,\Msunyr. The radiative-transfer calculations were carried out with the line-blanketed non-LTE model-atmosphere code \cmfgen\ \citep{1998ApJ...496..407H, 2005A&A...437..667D}, which solves the radiation-transfer equation for expanding media in the co-moving frame, assuming spherical symmetry and steady-state, and under the constraints set by the radiative-equilibrium and statistical-equilibrium equations. It treats line and continuum processes and regions of both small and high velocities (small and high velocities relative to the thermal velocity of ions and electrons). Hence, it can solve the radiative-transfer problem for both O stars, in which the formation regions of the lines and continuum extend from the hydrostatic layers out to the supersonic regions of the wind, and Wolf-Rayet stars, in which lines and continuum both originate in regions of the wind that may have reached half its asymptotic velocity.

\begin{figure}
 \centering
\includegraphics[width=8.5cm,angle=0]{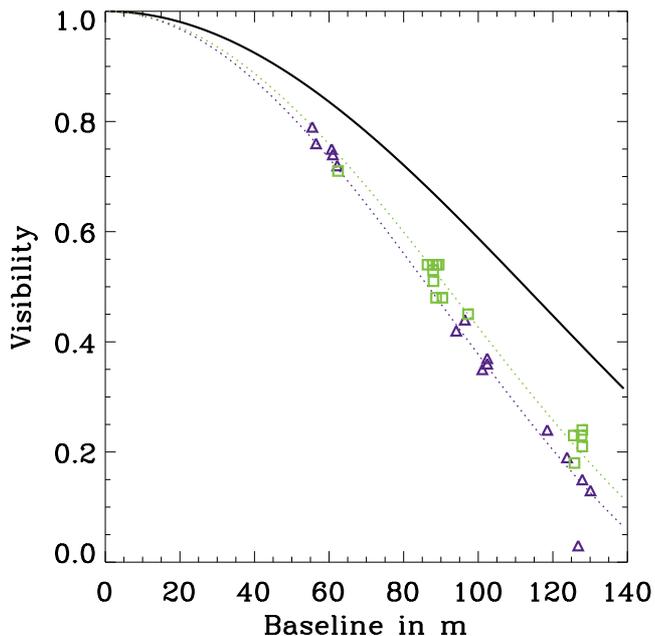}
 \caption[]{Depth of the differential visibility curves for \object{Rigel} referenced to the uniform-disk diameter curve, corresponding to 2.76\,mas. The {\sc Amber} measurements of the differential visibility obtained in the core of the line are indicated by triangles (2006-2007 campaign) and squares (2009-2010 campaign). A fit of theses values by two uniform-disk diameter curves provides with angular diameters of $3.58\pm0.04$\,mas and $3.45\pm0.04$\,mas, for 2006-2007 and 2009-2010, respectively.
\label{fig:CMFGEN_Visibilities}} 
\end{figure} 
Figure~\ref{fig:CMFGEN_Visibilities} shows the computed theoretical visibility curves for \object{Rigel} across the \Brg\ line with a mass-loss rate of $8 \times 10^{-7}$\,\Msunyr. The corresponding differential visibility profiles are overlaid with the observed differential visibilities in Figs.~\ref{fig:Rigel_vis2006}, \ref{fig:Rigel_vis2009} and \ref{fig:Rigel_vis2010} and in general provide a good match to the observations. However, the observed differential visibility profiles do show variations with respect to the steady-state model profiles either as red- or blue-shift of the visibility profile or variation of its width. 
\begin{table}[h!]
  \caption{\cmfgen parameters for Rigel.  }
    \label{tab:cmfgen}
  \centering
  \begin{tabular}{lc}
    \hline\hline
    	Parameter & Value \\
    \hline
	Luminosity \Lstar 				& $2.79 \times 10^5$\,\Lsun\\
	Terminal wind velocity \vinfty 		& $300$\,\kms  \\
	Stellar radius \Rstar 				& $115$\,\Rsun \\
	Mass loss \Mdot 				& $1, 2, 4, 6, 7, 8, 9, 10 \times 10^{-7}$\,\Msunyr\\
	Effective Temperature T$_{\rm eff}$ 	& $12000$\,K\\
	Filling factor & 10\%\\
	Composition & solar \\
     \hline
\end{tabular}
\end{table}
By fitting Gaussian curves on the differential visibilities and the spectra, we estimated the position of the core of the visibility in the line, and the FWHM, in \kms. From December 2006 to March 2007, the core is blueshifted from about $-20$\,\kms, moving to $0$\,\kms\ in March. The FWHM remains fairly constant, about $100\pm10\,$\kms. During that period of time, the spectral changes in the \Brg\ line are more noticeable. We have an absorption peaking at $40$\,\kms\ in December 2006. In March 2007, the line is centered on the reference value, and exhibits symmetrical dips. The 2009-2010 period is much more quiet compared to 2006-2007. The \Brg\ visibility signal remains well-centered at zero velocity. This variability is indicative of large-scale inhomogeneities in the line forming region but will not be discussed any further here.

The visibilities measured in the core of the \Brg\ line during 2009-2010 are well described by a uniform-disk diameter of $2.76$\,mas increased by $25$\%, corresponding to $3.45\pm0.04$\,mas. For the 2006-2007 data set a uniform-disk diameter of $3.58\pm0.04$\,mas is measured. Both determinations correspond to an extension of the Br$\gamma$ line forming region of $\sim$\,1.25\,R$_{\star}$. Given the number of measurements and the error bars, the 4\% difference in the extent of the \Brg\ line-forming region between the two periods is significant.

Using \cmfgen\ and varying only the mass-loss rate parameter, we performed a fit of the differential visibilities for each triplet of baselines that are shown in Fig.~\ref{fig:Rigel_vis2006}, \ref{fig:Rigel_vis2009} and \ref{fig:Rigel_vis2010}. Bearing in mind the limited statistics, such an analysis yields a mass-loss rate of $9.4\pm0.9 \times 10^{-7}$\Msunyr\ and $7.6\pm1.1 \times 10^{-7}$\Msunyr\ for the 2006-2007 and 2009-2010 campaigns, respectively. This implies variations in mass-loss rate at a level of 20-25\% over a period of several months. As discussed in Sect.~\ref{sec:continuum}, a change of the continuum diameter of Rigel between the two epochs is probable and would affect this analysis. 
Hence, the estimated variations in mass-loss rate could even be under-estimated when taking into account a change of the diameter of Rigel as a consequence of a variation in mass-loss rate.

\begin{figure}
\includegraphics[width=9cm,angle=0]{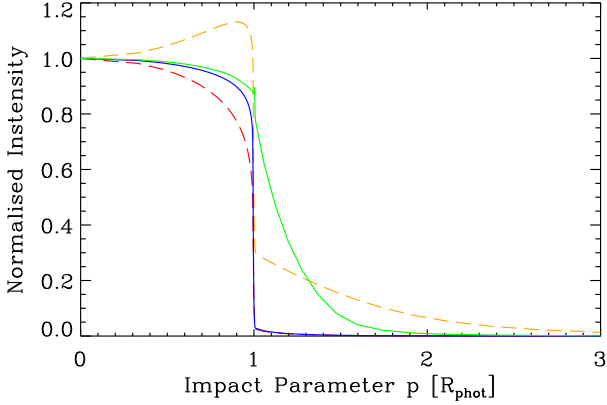}
\caption[]{A \cmfgen\ comparison between the normalized intensity profiles I(p) of \object{Rigel} in the \Ha\ line and its nearby continuum (dashed yellow and red lines, respectively), and the \Brg\ line and it nearby continuum (solid green and blue lines, respectively) for the model 
with \Mdot\,$=8 \times 10^{-7}$\Msunyr.
}\label{fig:CMFGEN}
\end{figure}

Figure~\ref{fig:CMFGEN} illustrates the variation of the intensity as a function of impact parameter and for different selected wavelengths ---
the model used has a mass-loss rate of $8 \times 10^{-7}$\,\Msunyr. 
This model predicts a larger uniform-disk diameter for H$\alpha$ than for Br$\gamma$. H$\alpha$ can probe structures up to 3 stellar radii while the Br$\gamma$ perturbations mostly originate at most at 1.5 stellar radii. Moreover, one expects the Br$\gamma$ perturbations to be brighter close to the photosphere because the  H$\alpha$ emitting region
is more extended.  This difference must be kept in mind in the frame of the differential phase/photocenter analysis performed in Sect.~\ref{sec:phase}. Using optical interferometric observations of \object{Rigel}, \citet{2010A&A...521A...5C} inferred a uniform-disk diameter of the H$\alpha$ line forming region of $4.2$\,mas, equivalent to $\sim1.5$\,R$_*$. However, the corresponding fit of the differential visibilities of H$\alpha$ required a much lower mass-loss rate of $1.5 \times 10^{-7}$\,\Msunyr\ than required for Br$\gamma$ in this study.  Further work on the \cmfgen\ model, out of the scope of the present paper, is needed to solve this inconsistency.  
\section{Differential phases and circumstellar structures}
\label{sec:phase}
%
Figure~\ref{fig:Rigel_phase2006} and Figs.~\ref{fig:Rigel_phase2009} and \ref{fig:Rigel_phase2010} show in the first three columns corresponding to the three available baselines the measured differential phases across the Br$\gamma$ line in the 2006-2007 and 2009-2010 data set, respectively.
%
In some differential phase measurements of \object{Rigel} a statistically significant, non-zero signal is detected across the Br$\gamma$ line. The signal is highly complex (and noisy) across the line and reaches from typical values of $\pm2$\,degrees to maximum values of $-6$ to $+8$\,degrees. 
This observation of a non-zero interferometric signal in the differential phases across the Br$\gamma$ line is indicative of asymmetry of the line forming region and therefore the circumstellar envelope. Yet, the signal is also very weak, suggesting that the perturbation can only be considered as a second-order effect. As shown in Fig.\ref{fig:Rigel_tsBrg2006} and Fig.\ref{fig:Rigel_tsBrg2009}, this corresponds also to moderate changes in the Br$\gamma$ line, never exceeding 10\% of the line profile.
The strongest differential phase signal can often be characterised by so-called 'S-shape' profiles (cf. e.g. Fig.~\ref{fig:Rigel_phase2009} at MJD\,55238) that are commonly interpreted as signatures of {\it rotating} circumstellar structures \citep[see for instance][]{2013LNP...857..149M}.  

To further explore the nature and characteristics of \object{Rigel}'s circumstellar structures and their temporal variation in absence of a detailed model that could be inverted to fit the measured differential phase, the
measured differential phases $\phi$ are converted into differential astrometric shifts $p$ using the well-known relation for marginally resolved interferometric observations \citep{2003A&A...400..795L, 1995A&AS..109..389C} 
\begin{equation}
\label{eqn:astrometric_shift}
p=-\frac{\phi}{2\pi}\cdot\frac{\lambda}{B}
\end{equation}

where $B$ is the projected baseline and $\lambda$ is the wavelength of the spectral channel. 
$p$ represents the projection in the baseline direction of the estimated 2D photocenter in the plane of the sky $\vec{p}$. With a triplet of baselines provided by a single observation with {\sc Amber} the astrometric solution (i.e., a single 2D vector $\vec{p}$) can be estimated reliably using a robust least-square minimisation scheme. Furthermore, this global fit is linear and can be performed at each wavelength through the Br$\gamma$ line. For a recent discussion on the relation between differential  phases and photocenter shifts see \citet{2012ASPC..464...15M}.

However, Eqn.~\ref{eqn:astrometric_shift} is formally only applicable to unresolved or marginally resolved objects with visibilities close to 1. Therefore, the limitations of applying this method to more extended and resolved objects like \object{Rigel} must be carefully considered. With the longest baselines, the Br$\gamma$ line-forming region of \object{Rigel} is well resolved with observed visibilities in the range of $0.1$ to $0.3$ and hence the astrometric signal is biased. 
Such a bias is difficult to evaluate quantitatively without a good knowledge of the geometry of the source. Qualitatively, the bias results in a distortion of the reconstructed patterns. This distortion can be considered as a systematic bias in the 2006-2007 dataset because all observations have been obtained with the same telescope configuration, i.e., the UT1-3-4 triplet (cf. Tab.~\ref{tab:log_obsNIR}).

The reconstructed photocenter shifts as a function of spectral channel across the Br$\gamma$ line of the \object{Rigel} data sets are shown in the right-most column of Fig.~\ref{fig:Rigel_phase2006} for the 2006-2007 dataset and in Figs.~\ref{fig:Rigel_phase2009} and \ref{fig:Rigel_phase2010} for the 2009-2010 dataset. 
The black diamonds represent the continuum points measured outside the velocity intervals $\pm$40 \kms\ centered on the line, the coloured diamonds represent the different wavelengths across the line.

The cloud of black diamonds in the center of the figures are the continuum points that provide a good insight on the level of the formal error bar due to the statistical variations of the differential phases. 
Well-structured photocenter shifts for the spectral channels within the Br$\gamma$ line are observed at several epochs of the 2006-2007 and 2009-2010 campaigns. The patterns exhibit similar shapes, whether in the form of a weakly-curved single-sided arm connected to the star or in the form of loops indicative of a more complex velocity structure. 
The measured characteristics of the observed structures are summarized in Tab.~\ref{tab:photocenter}.
\begin{table}[h!]
  \caption{Position angles and extensions of the structures measured in the photocenter shifts in Figs.~\ref{fig:Rigel_phase2006}, \ref{fig:Rigel_phase2009} and \ref{fig:Rigel_phase2010}. }
    \label{tab:photocenter}
  \centering
  \begin{tabular}{lrrl}
    \hline\hline
    \multicolumn{4}{c}{2006-2007 campaign}\\
    \hline
    	MJD & PA [\deg] & L [mas] & Comment  \\
    \hline
	54099.104 &  145   & 0.05 & small loop\\
	54107.174 &    45   & 0.12 & single-sided arm\\
	54108.058 &    50   & 0.12 & single-sided arm\\
	54137.055 &  135   & 0.12 & narrow loop\\
	54168.008 &  $>135,<260$ & 0.06, 0.08  & wide loop\\
    \hline
        \multicolumn{4}{c}{2009-2010 campaign}\\
   \hline
   	MJD & PA [\deg] & L [mas] & Comment  \\
    \hline
 	55139.263 &  $>30,<160$ & 0.06, 0.05 & wide loop\\
	55152.258 &  $<30,>200$ & 0.06, 0.12 & wide loop\\
	55189.175 &  --- & $<0.01$ & no structure\\
	55217.061 &  315 & 0.04 & small loop\\
	55238.039 &  0,205 & $0.05, 0.10$ & wide loop\\
	55267.037 & --- & --- & could not be determined \\
	55275.084 & --- & --- & undefined structures\\
	55299.986 & 200 & 0.07  & single-sided arm\\
   \hline
 \end{tabular}
\end{table}

Before discussing the observed extended structures it is worth to note that at MJD\,55189 {\it no} photocenter shifts larger than $0.01$\,mas corresponding to 0.4\% of the stellar radius were observed indicating a mostly undisturbed, spherical-symmetric envelope, i.e., a rather uncommon state of quiescence for \object{Rigel}. Unfortunately, no optical spectrum was recorded on the same night but the H$\alpha$ spectrum closest in time, i.e., on MJD\,55178 displays a classical P-Cygni line profile (cf.~Fig.~\ref{fig:Rigel_tsHaBrgdiff2009}) with red-shifted emission and blue-shifted absorption, compatible with a spherical-symmetric stellar wind. 
It should be noted here that the observed photocenter shifts never exceed $0.12$\,mas corresponding to $1.04$\,\Rstar, i.e., the observed structures to which the \Brg\ lines is sensitive and which are described in the following are located very close to the star.

The first one-armed structure observed in the photocenter shifts in 2006-2007 on MJD\,54107 and 54108 appears rather stable from one day to the next with a possible (counter-clock wise) increase of the position angle of a few degrees. Some 30\,days later (MDJ\,54137) the pattern seems to have rotated counter-clockwise by about 90\,degrees. The pattern has also curved and has the appearance of a loop. Most of the signal is found in the blue side of the line, evolving from blue to red. 
Observations performed another 30 days later (MDJ\,54168) are of lower quality as witnessed by the asymmetry of the continuum cloud of points (black diamonds). Yet, a significant differential signal is observed near the zero velocity, giving rise to a weak but significant arm whose angle seems to fit consistently into the counter-clockwise rotation observed before.

The timescale of 30\,days for a quarter of a full rotation is consistent with the estimated upper limit for the stellar rotation period of 107\, days \citep{1996A&A...305..887K}.

No such consistent pattern evolution is observed in the 2009-2010 campaign. At MJD\,55139, a complex, loopy pattern is observed. Unfortunately, at MJD\,55152, only one good quality differential phase was obtained, hampering the photocenter interpretation. At MJD\,55217, a small loop pattern is observed that develops at MJD\,55238 into an extended loop pattern that seems to extend in two almost opposite directions and therefore involves a large circumstellar volume.

The  largest photocenter shift and the strongest observed S-shaped signal in the differential phases of the complete available data set was observed at MJD\,55238. This measurement connects with the strongest observed feature in the spectral time series in H$\alpha$, i.e., a comparatively weak HVA event becoming visible about 30 days later at MJD\,55264. 
Admittedly, the sequential appearance of the two features could be pure coincidence which can not be ruled out due to the lack of temporal sampling of the interferometric data set. However, since the time lag of 25\,days can possibly be identified with one quarter of the estimated rotation period of \object{Rigel} \citep{1996A&A...305..887K}, it is possible that we observe the rotation of an extended circumstellar structure from next to the star (observable in the differential phase of Br$\gamma$ through a photocenter shift) into the line of sight in front of the star (observable as blue-shifted absorption in the H$\alpha$ line profile).   
\section{Results and Discussion}
\label{sec:discussion}
We interpret the \Ha\ and \Brg\ spectral variability as variable emission superimposed on expanding wind profiles. The 2006-2007 visibilities data set (in comparison with 2009-2010 data set) indicates an increased size of the photosphere and the \Brg\ forming region. We may speculate that such an increase linked to the large mass-ejection event seen at MJD\,54048 (high-velocity absorption HVA in line of sight extending to $-400$\,\kms). The \cmfgen\ modelling show that a mass-loss change of about 20\% between the two epochs can explain the variation of the differential visibilities (see Fig.~\ref{fig:Rigel_vis2006}, \ref{fig:Rigel_vis2009} and \ref{fig:Rigel_vis2010}).
We detected many occurrences in the differential phases in \Brg\ of some 'S-shaped' signals indicative of rotating circumstellar material that also causes the spectral variation. Linking these differential phases to photocenter shifts by a linear relationship provides some evidence that these structures are extended, reminiscent of loops. The strongest S-shape signal is observed in 2009-2010 data set around MJD\,55240. This event was not preceded between MJD\,55265 and MJD\,55280 (end of data set) by any strong features in \Ha\ followed by a strong absorption event crossing the \Ha\ line profile from $-250$ to $+150$\,\kms\ nor by any strong features in the interferometric signal at that time. The \Brg\ line-forming region being closer to the star than \Ha, it is probable that the detection of the interferometric signal occurred shortly after the material ejection or wind perturbation. 

The interferometric monitoring of such events shows that they occur at a low rate of 1-2 per year. The differential phases show that this ejection is {\it local}, from a defined location at the star surface. Elaborated hydrodynamical 2D and 3D models of CIRs were shown in \citet{2004A&A...423..693D} and \citet{2002A&A...395..209D}.  Theoretical expectations for interferometric observations of such a perturbed hot star outflow were proposed and it was shown that strong spectrally dispersed signal of the same nature as those determined from spectroscopy were present in the differential phases. Applying the same linear photocenter-shift relation used in this paper (and therefore being subject to similar biases) the theoretical signatures have the appearance of a loop or arm that rotates around the star. In their Figs.~11 and 12, \citet{2002A&A...395..209D} showed the cumulative signal over a rotation period and its dependence on the star inclination. In the monitoring presented here, we are far  from having the time and spatial coverage to compare to these plots, but one can note some encouraging resemblance. First the level of the signal, about 0.05 to 0.1\,mas is comparable to the level shown in the theoretical study. Second, the appearance of the differential signal and its conversion into a photocenter shift is also similar. In the theoretical study of \citet{2002A&A...395..209D}, two perturbations were artificially included as the origin of two CIRs at each side of the star but only one pattern is observed at any one time around \object{Rigel}.

\section{Conclusions}
\label{sec:conclusions}

We have reported on the first optical interferometric campaign aiming at studying the activity observed around the bright blue 
supergiant star \object{Rigel}. The organisation of a long-term monitoring is challenging and imposes stringent constraints on the {\sc Vlti} infrastructure to obtain as regularly as possible observations with a configuration of telescopes that optimally would be close to an equilateral triangle. In both campaigns, this condition could not be fulfilled due to the complexity of the telescope relocation management for this open facility. Furthermore, a spectral resolution of $R=12000$ is required to resolve the \Brg\ line due to the small \vsini\ of \object{Rigel}. Only {\sc Amber/Vlti} in the southern hemisphere and {\sc Vega/Chara} in the northern hemisphere can currently perform such a challenging and demanding temporal monitoring. 

The results obtained are promising. It has been shown that the \Brg\ line-forming region is well resolved by the interferometer. \cmfgen\ modelling implies a mass-loss rate at least a factor two larger than in the very similar model published by \citet{2010A&A...521A...5C} to account for the {\sc Vega/Chara} interferometric observations in the \Ha\ line. It is out of the scope of this paper to further investigate the origin of such a discrepancy. However, these observations illustrate the need for a global approach to obtain robust models of B supergiant photospheres and winds. The extent of the \Brg\ line-forming region was established at about $1.25$\,\Rstar\ but also found to be variable. This variability translates into mass-loss variations of at least 20\% on a timescale of one year. 

Strong activity is observed in the differential and closure phase, although at a low level. At some periods, no phase signal was observed at all. This confirms that the observed circumstellar activity is better understood in the context of second-order perturbations of an underlying spherical wind whose properties can be well reproduced by a 1D radiative transfer code. 

The phase signal is spectrally extended, implying a large physical extent of the perturbation. The observed temporal variations suggest slowly evolving structures. Such a behaviour cannot be explained by the random emission of large scale clumps. Instead, the differential phases resemble the signal expected from theoretical models of CIRs. 
The well-structured signal implies that the activity around \object{Rigel} is very low, with some moderate eruptions occurring at a month to years time scale. The detected structures can be followed in the \Brg\ line over 2-4 months, i.e., typically on a rotation time scale.

These conclusions are in line with results from numerous intensive spectroscopic monitoring campaigns that have to date been performed on \object{Rigel} and other massive hot supergiants.

%
\begin{acknowledgements}
We thank J.B. LeBouquin and Ph. Berio for their help with the photocenter inversion. This publication is supported as a project of the Nordrhein-Westf\"alische
Akademie der Wissenschaften und der K\"unste in the framework of the academy
program by the Federal Republic of Germany and the state
Nordrhein-Westfalen.
\end{acknowledgements}
%
%

%
\clearpage
%
\appendix
\section{Observation logs}
\begin{table*}[h!]
  \caption{Observation log of the {\sc Feros} and {\sc Beso} observations of \object{Rigel}. 
  Column $n$ indicates the number of spectra that have been combined from the respective night.
  }
  \label{tab:log_obsOptical}
  \centering    
  \begin{tabular}[h]{rllr|rllr} \hline\hline
  \multicolumn{8}{c}{2006-2007 campaign with {\sc Feros}} \\
  \hline
  Spectrum & Date       & MJD            &$n$ & Spectrum & Date       & MJD            &$n$  \\
  \hline
 1 & 2006-10-03 & 54011.36031973 & 6  & 11 & 2006-12-27 & 54096.31590096 & 5    \\ 
 2 & 2006-10-07 & 54015.28385735 & 6  & 12 & 2007-01-01 & 54101.16567652 & 36   \\ 
 3 & 2006-10-11 & 54019.36730933 & 3  & 13 & 2007-01-17 & 54117.14069429 & 3    \\ 
 4 & 2006-10-29 & 54037.31208022 & 20 & 14 & 2007-02-05 & 54136.05338341 & 10   \\ 
 5 & 2006-12-02 & 54071.25872330 & 6  & 15 & 2007-02-11 & 54142.10571373 & 6    \\ 
 6 & 2006-12-05 & 54074.30587822 & 5  & 16 & 2007-02-16 & 54147.06192037 & 9    \\ 
 7 & 2006-12-08 & 54077.19925441 & 13 & 17 & 2007-02-22 & 54153.09321235 & 10   \\ 
 8 & 2006-12-14 & 54083.17843854 & 10 & 18 & 2007-03-24 & 54183.99706054 & 6    \\ 
 9 & 2006-12-18 & 54087.28252888 & 6  & 19 & 2007-03-30 & 54189.02635471 & 3    \\ 
10 & 2006-12-22 & 54091.14807436 & 11 & 20 & 2007-03-31 & 54191.00199574 & 3    \\ 
  \hline
  \multicolumn{7}{c}{2009-2010 campaign with {\sc Beso}} \\
  \hline
  Spectrum  & Date       & MJD            &  &  Spectrum & Date       & MJD            &  \\
  \hline
 1 & 2009-10-06 & 55110.33170140  &  &      41 & 2010-02-23 & 55250.00540509 &  \\
 2 & 2009-10-17 & 55121.23127315 &  &      42 & 2010-02-24 & 55251.02425926 &  \\
 3 & 2009-10-20 & 55124.26278935 &  &      43 & 2010-02-25 & 55252.09684028 &  \\
 4 & 2009-10-22 & 55126.24055556 &  &      44 & 2010-02-26 & 55253.04282407 &  \\
 5 & 2009-10-24 & 55128.23674768 &  &      45 & 2010-02-28 & 55255.02221065 &  \\
 6 & 2009-10-28 & 55132.21540509 &  &      46 & 2010-03-01 & 55256.00712963 &  \\
 7 & 2009-10-30 & 55134.27407408 &  &      47 & 2010-03-03 & 55258.00199074 &  \\
 8 & 2009-11-02 & 55137.19428241 &  &      48 & 2010-03-05 & 55260.99156253 &  \\
 9 & 2009-11-04 & 55139.18910879 &  &      49 & 2010-03-06 & 55261.98248839 &  \\
10 & 2009-11-13 & 55148.16081018 &  &      50 & 2010-03-07 & 55262.98533567 &  \\
11 & 2009-11-15 & 55150.14854167 &  &      51 & 2010-03-08 & 55263.98876158 &   \\
12 & 2009-11-17 & 55152.20844907 &  &      52 & 2010-03-09 & 55264.98519675 &   \\
13 & 2009-11-19 & 55154.20519676 &  &      53 & 2010-03-10 & 55265.97914354 &   \\
14 & 2009-11-21 & 55156.16405093 &  &      54 & 2010-03-13 & 55268.00694444 &   \\
15 & 2009-12-01 & 55166.28560185 &  &      55 & 2010-03-14 & 55269.02199074 &   \\
16 & 2009-12-05 & 55170.08598379 &  &      56 & 2010-03-14 & 55269.97369210  &   \\
17 & 2009-12-09 & 55174.17329862 &  &      57 & 2010-03-15 & 55270.99855328 &   \\
18 & 2009-12-11 & 55176.15767361 &  &      58 & 2010-03-16 & 55271.98635419 &   \\
19 & 2009-12-13 & 55178.09023148 &  &      59 & 2010-03-17 & 55272.97731479 &   \\
20 & 2010-01-12 & 55208.06005787 &  &      60 & 2010-03-18 & 55273.98398145 &   \\
21 & 2010-01-12 & 55208.06197916 &  &      61 & 2010-03-20 & 55275.01642361 &  \\
22 & 2010-01-14 & 55210.05552083 &  &      62 & 2010-03-22 & 55277.05741898 &  \\
23 & 2010-01-16 & 55212.05674768 &  &      63 & 2010-03-22 & 55277.99923611 &  \\
24 & 2010-01-18 & 55214.05457176 &  &      64 & 2010-03-23 & 55278.97440974 &  \\
25 & 2010-01-18 & 55214.05793981 &  &      65 & 2010-03-25 & 55280.05094907 &  \\
26 & 2010-01-22 & 55218.10179398 &  &      66 & 2010-03-25 & 55280.96922453 &  \\
27 & 2010-01-25 & 55221.02671296 &  &      67 & 2010-03-26 & 55281.97070599 &  \\
28 & 2010-01-29 & 55225.14851852 &  &      68 & 2010-03-27 & 55282.96931712 &  \\
29 & 2010-01-31 & 55227.12858796 &  &      69 & 2010-03-28 & 55283.96629628 &  \\
30 & 2010-02-05 & 55232.02315972 &  &      70 & 2010-03-29 & 55284.97012734 &  \\
31 & 2010-02-08 & 55235.00275463 &  &      71 & 2010-03-30 & 55285.98187502 &  \\
32 & 2010-02-10 & 55237.06417824 &  &      72 & 2010-03-31 & 55286.98141201 &  \\
33 & 2010-02-12 & 55239.10770833 &  &      73 & 2010-04-01 & 55287.97542826 &  \\
34 & 2010-02-14 & 55241.11258101 &  &      74 & 2010-04-02 & 55288.97332176 &  \\
35 & 2010-02-16 & 55243.01336806 &  &      75 & 2010-04-03 & 55289.97159719 &  \\
36 & 2010-02-18 & 55245.03077546 &  &      76 & 2010-04-04 & 55290.96679401 &  \\
37 & 2010-02-19 & 55246.00457176 &  &      77 & 2010-04-05 & 55291.97576388 &  \\
38 & 2010-02-19 & 55246.99271989 &  &      78 & 2010-04-07 & 55293.00043981 &  \\
39 & 2010-02-21 & 55248.00678241 &  &      79 & 2010-04-07 & 55293.99209491 &  \\
40 & 2010-02-22 & 55249.04751157 &  &         &            &                &  \\
 \hline
\end{tabular}
\end{table*}
\clearpage
\begin{table*}[h!]
  \caption{Observation log of {\sc Amber}. The projected baselines at the dates of observation are provided in length and position angle (PA).}
    \label{tab:log_obsNIR}
  \centering
  \begin{tabular}{rlllrrrrrr}
    \hline\hline
    \multicolumn{10}{c}{2006-2007 campaign with {\sc Amber}}\\
    \hline
        Spectrum & Date & MJD & Stations & \multicolumn{3}{c}{Length [m]} & \multicolumn{3}{c}{PA [degrees]}  \\
    \hline
     1 & 2006-12-30 & 54099.104 & UT1--3--4 & 96.4& 62.2& 123.8 & 28.8& 108.3& 58.4 \\
     2 & 2007-01-07 & 54107.174 & UT1--3--4 & 102.4& 56.5& 127.9 & 38.8& 115.4& 64.2 \\
     3 & 2007-01-08 & 54108.058 & UT1--3--4 & 94.1& 61& 118.5 & 25.0& 107.6& 55.7 \\
     4 & 2007-02-05 & 54137.055 & UT1--3--4 & 101.1& 60.7& 130.1 & 36.0& 111.7& 62.9 \\
     5 & 2007-03-08 & 54168.008& UT1--3--4 & 102.4& 55.5& 126.8 & 39.1& 116.2& 64.3 \\
    \hline
        \multicolumn{10}{c}{2009-2010 campaign with {\sc Amber}}\\
   \hline
       Spectrum & Date & MJD & Stations & \multicolumn{3}{c}{Length [m]} & \multicolumn{3}{c}{PA [degrees]}  \\
    \hline
    1 & 2009-11-04  & 55139.263 & UT1--3--4 & 97.3& 62.4& 135.5 & 30.2& 108.7& 59.4 \\
    2 & 2009-11-04 & 55139.279 & UT1--3--4 & 98.4& 62.4& 127.5 & 31.9& 109.3& 60.4 \\
    3 & 2009-11-17 & 55152.258 & A0--K0--G1 & 87.9& 89.4& 127.7 & -152.0& -64.4& -107.0 \\
    4 & 2009-12-24 & 55189.175 & A0--K0--G1 & 88.8& 87.8& 127.8 & -150.0& -63.1& -107.0 \\
    5 & 2010-01-21 & 55217.061 & A0--K0--G1 & 86.4& 90.4& 125.7 & -155.0& -65.9& -109.0 \\
    6 & 2010-02-11 & 55238.039 & A0-K0-G1 & 88.7& 88.0& 127.9 & -150.0& -63.3& -107.0 \\
    7 & 2010-03-12 & 55267.037  & D0--H0--K0 & 56.5& 28.2& 84.8 & 74.8& 74.8& 74.8 \\
    8 & 2010-03-20 & 55275.084 & A0--K0--G1 & 86.0& 53.8& 78.9 & -145& -29& -107 \\
    9 & 2010-04-13 & 55299.986 & D0--I1--G1 & 53.5& 43.3& 55.4 & 114.5& -134& 161.3 \\
   \hline
 \end{tabular}
\end{table*}
\clearpage
\section{2006-2007 campaign}
 \begin{figure}[h!]
   \centering   \includegraphics[width=0.48\textwidth]{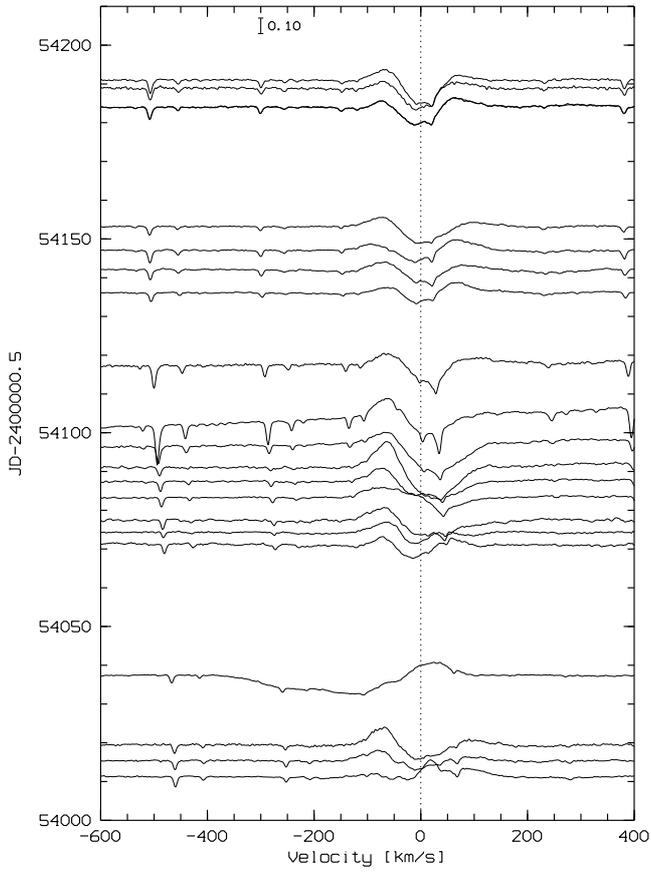}
   \caption{2006-2007 \Ha\ time series of \object{Rigel} obtained with {\sc Feros}.}
   \label{fig:Rigel_tsHa2006} 
 \end{figure}
 \begin{figure}[h!]
   \vspace*{1.3cm}
   \centering   \includegraphics[width=0.48\textwidth]{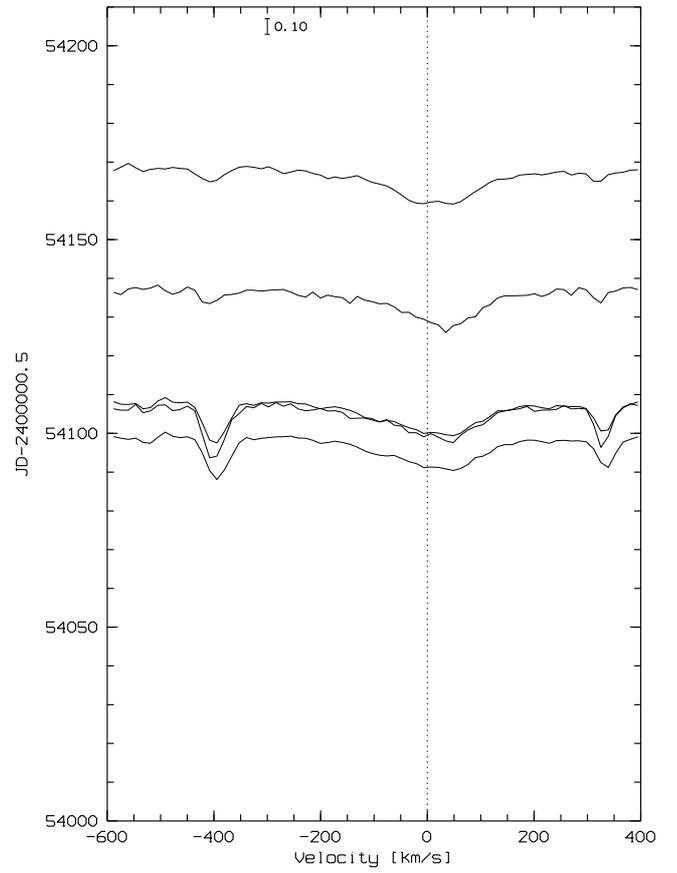}
   \caption{2006-2007 \Brg\ time series of \object{Rigel} obtained with {\sc Amber}.}
   \label{fig:Rigel_tsBrg2006} 
 \end{figure}
 \begin{figure*}[h!]
	\centering \includegraphics[width=0.98\textwidth]{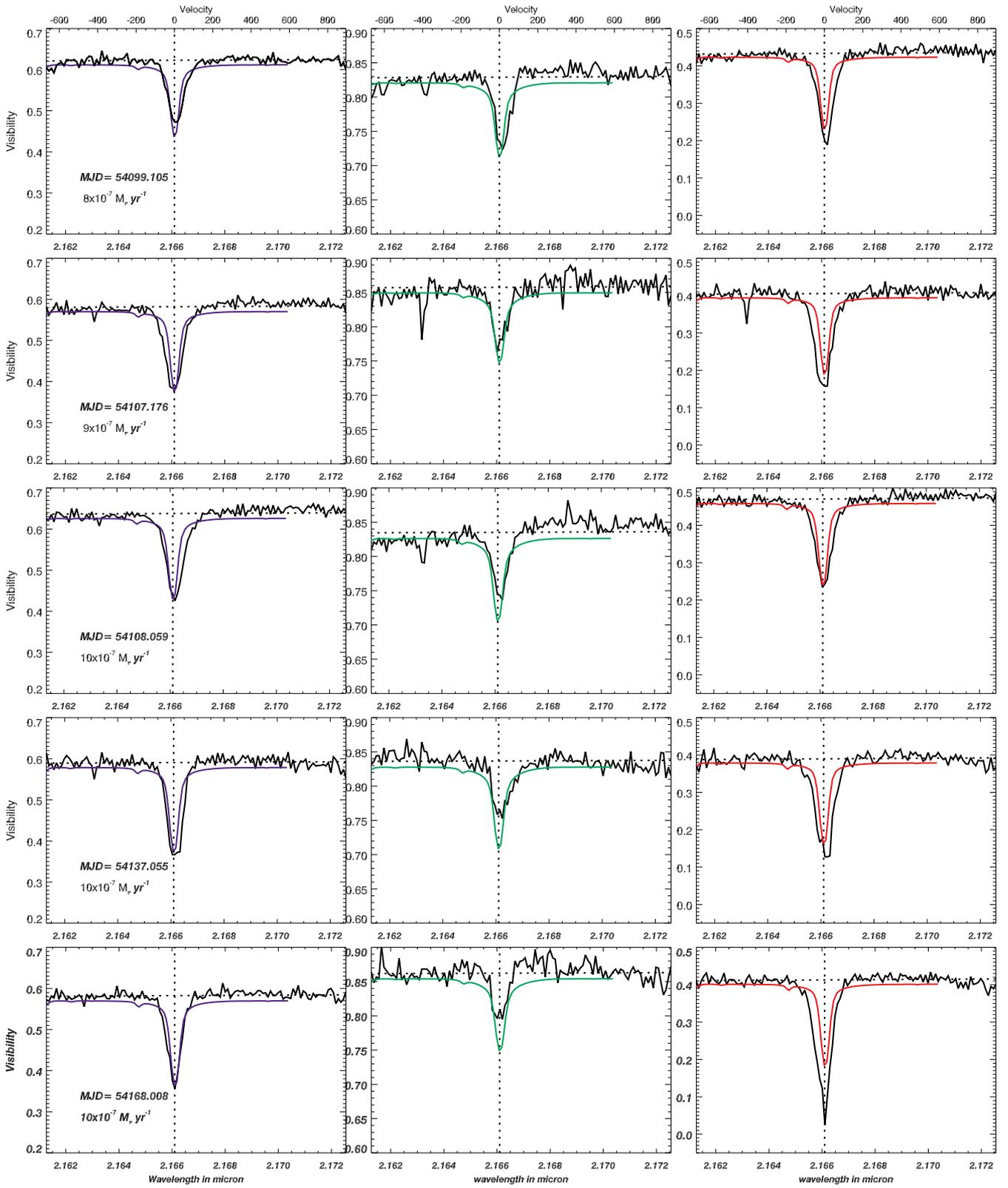}
   \caption{ Differential visibilities of \object{Rigel} obtained in 2006-2007 and put to the scale corresponding to an angular diameter of 2.75\,mas. The visibilities are compared with the ones computed from \cmfgen\ models tuning the mass-loss rate only.}
   \label{fig:Rigel_vis2006} 
 \end{figure*}
 \begin{figure*}[h!]
	\centering   \includegraphics[width=0.98\textwidth]{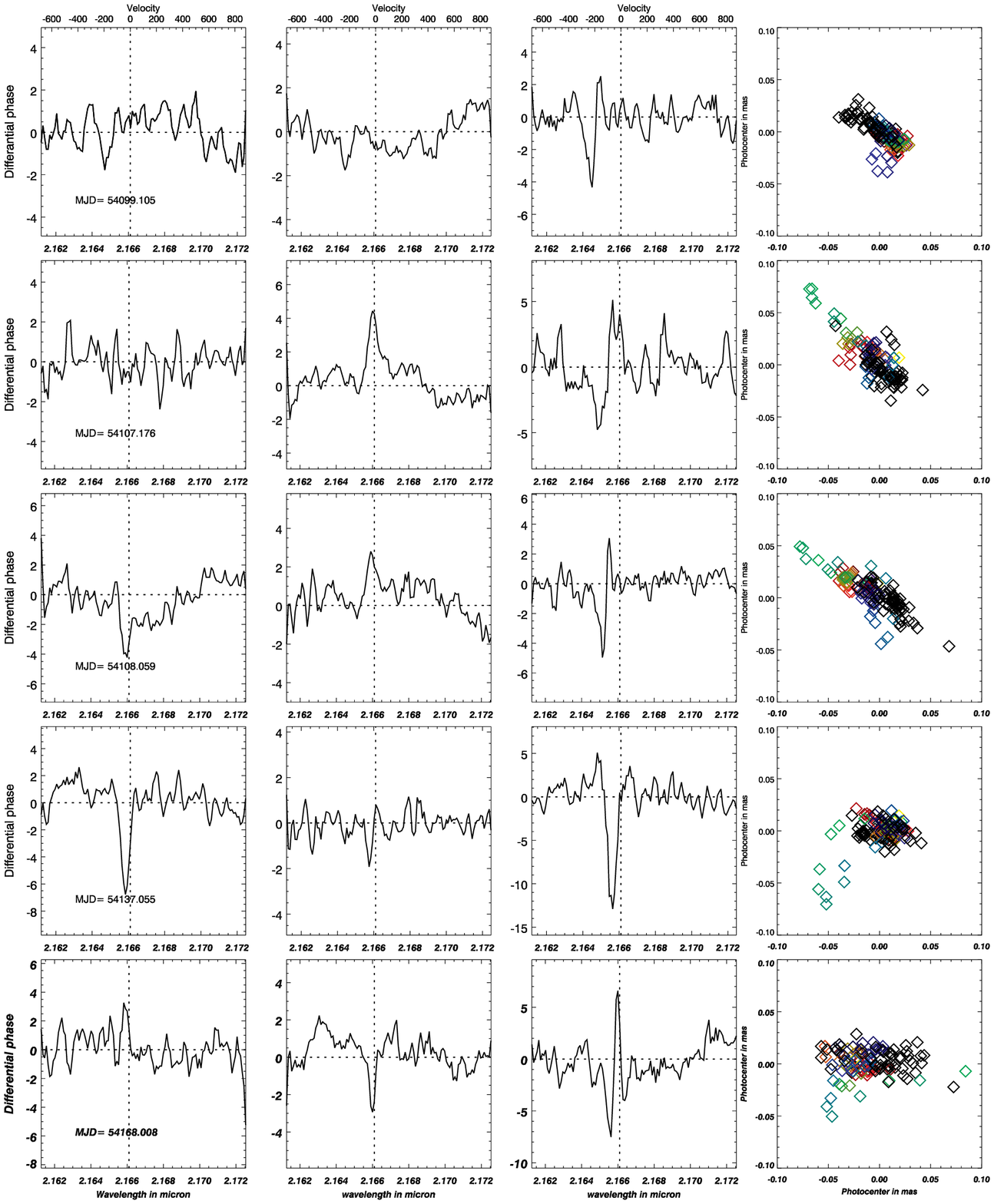} 
   \caption{Differential phases and photocenter shift of \object{Rigel} obtained in 2006-2007.}
   \label{fig:Rigel_phase2006} 
 \end{figure*} 
\clearpage 
\section{2009-2010 campaign}
 \begin{figure}[h!]
   \centering   \includegraphics[width=0.48\textwidth]{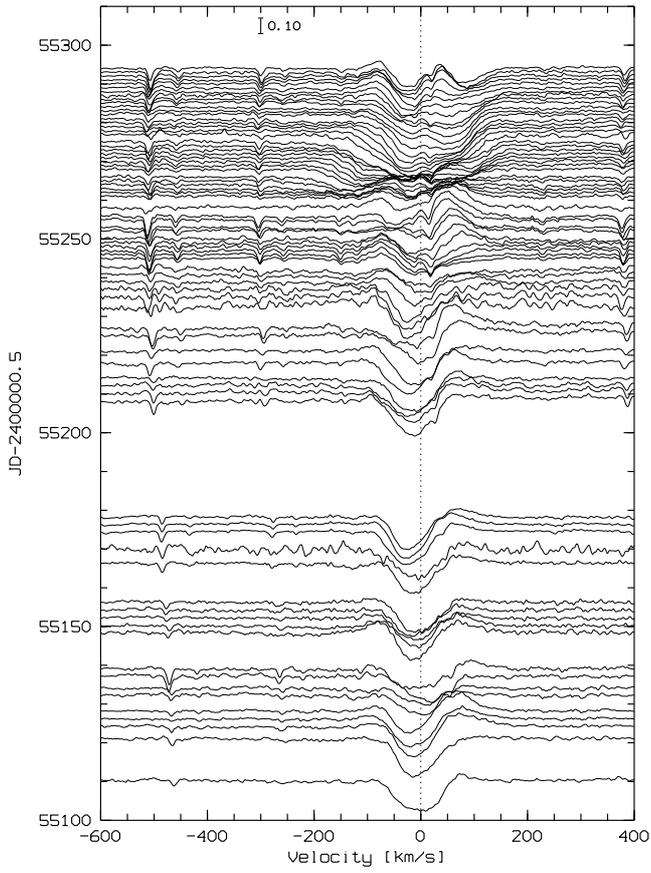}
   \caption{2009-2010 \Ha\ time series of \object{Rigel} obtained with {\sc Beso}.}
   \label{fig:Rigel_tsHa2009} 
 \end{figure}
 \begin{figure}[h!]
    \vspace*{1.3cm}
   \centering   \includegraphics[width=0.48\textwidth]{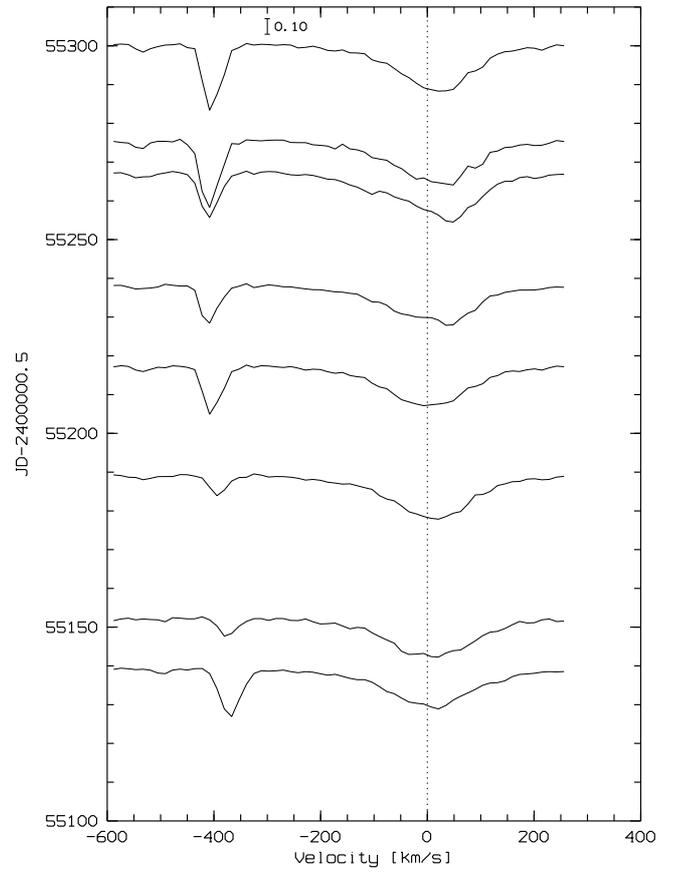}
   \caption{2009-2010 \Brg\ time series of \object{Rigel} obtained with {\sc Amber}.}
   \label{fig:Rigel_tsBrg2009} 
 \end{figure}
 \begin{figure*}[h!]
   \centering
	\includegraphics[width=0.98\textwidth]{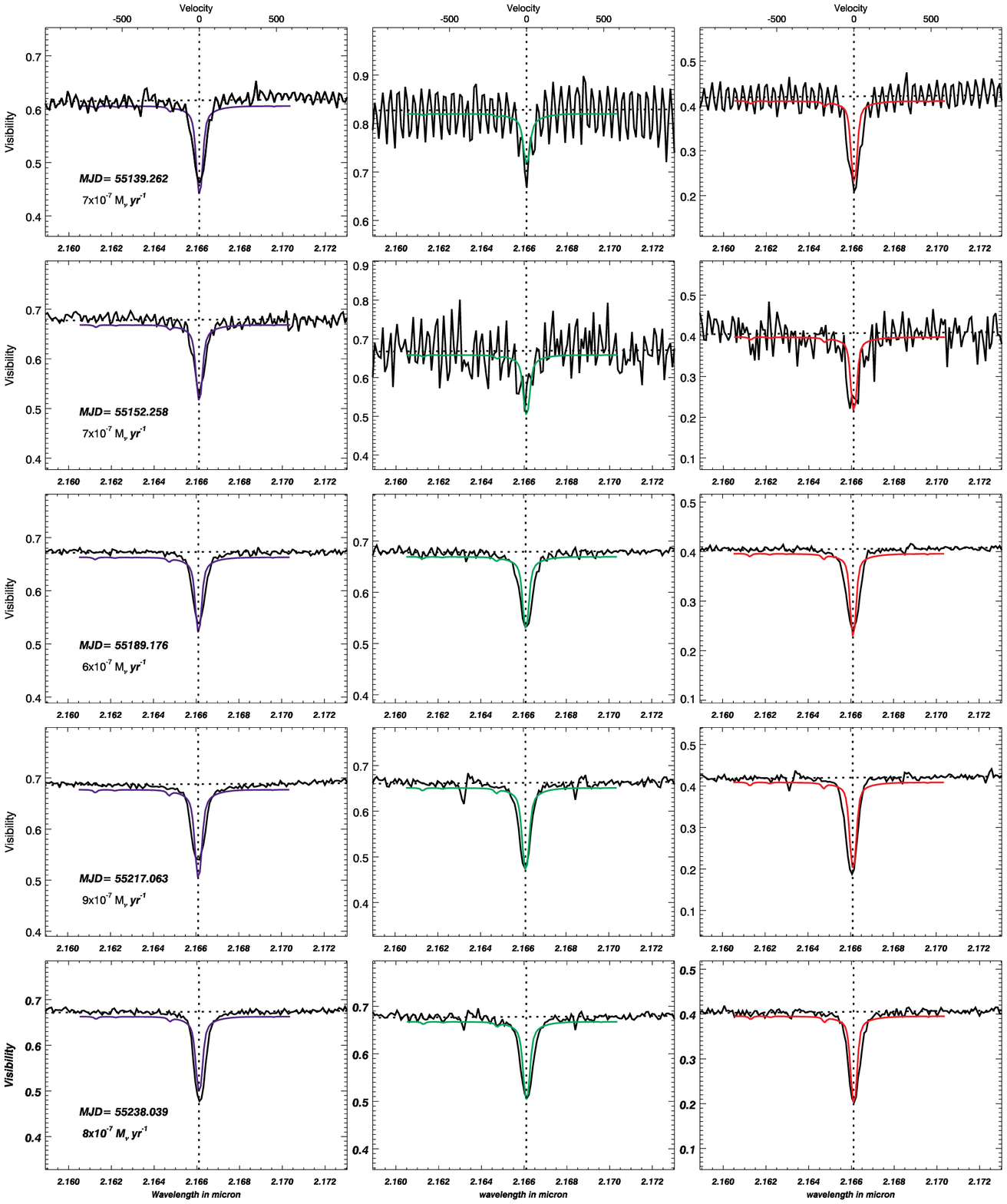}
   \caption{Differential visibilities of \object{Rigel} obtained obtained in 2009-2010 and put to the scale corresponding to an angular diameter of 2.75\,mas. The visibilities are compared with the ones computed from  \cmfgen\ models with varying mass-loss rate.}
   \label{fig:Rigel_vis2009} 
  \end{figure*}
  \begin{figure*}[h!]
   \centering
	\includegraphics[width=0.98\textwidth]{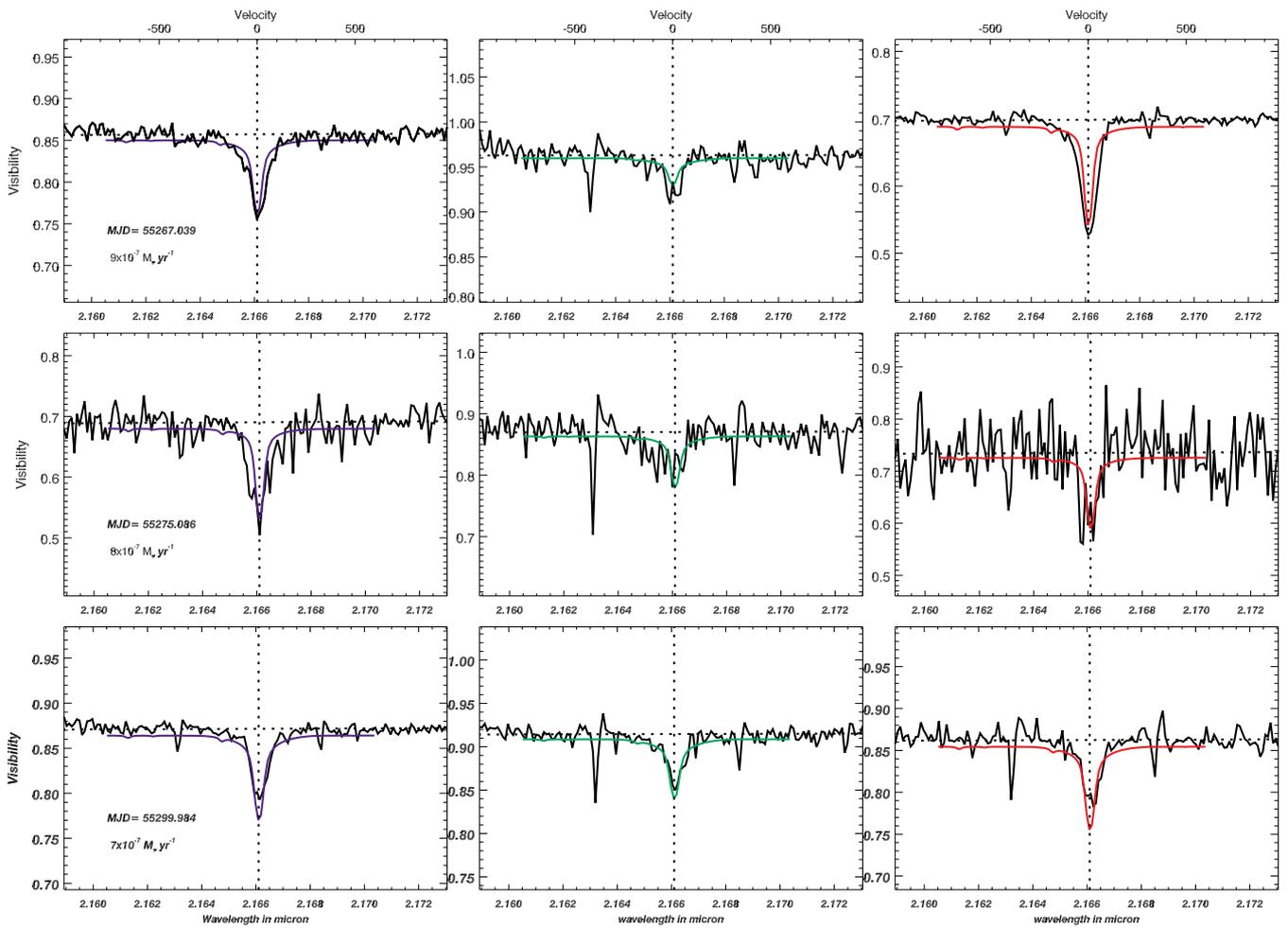}
   \caption{Differential visibilities of \object{Rigel} obtained in 2009-2010 (continued).}
   \label{fig:Rigel_vis2010} 
  \end{figure*}
 \begin{figure*}[h!]
   \centering
      \includegraphics[width=0.98\textwidth]{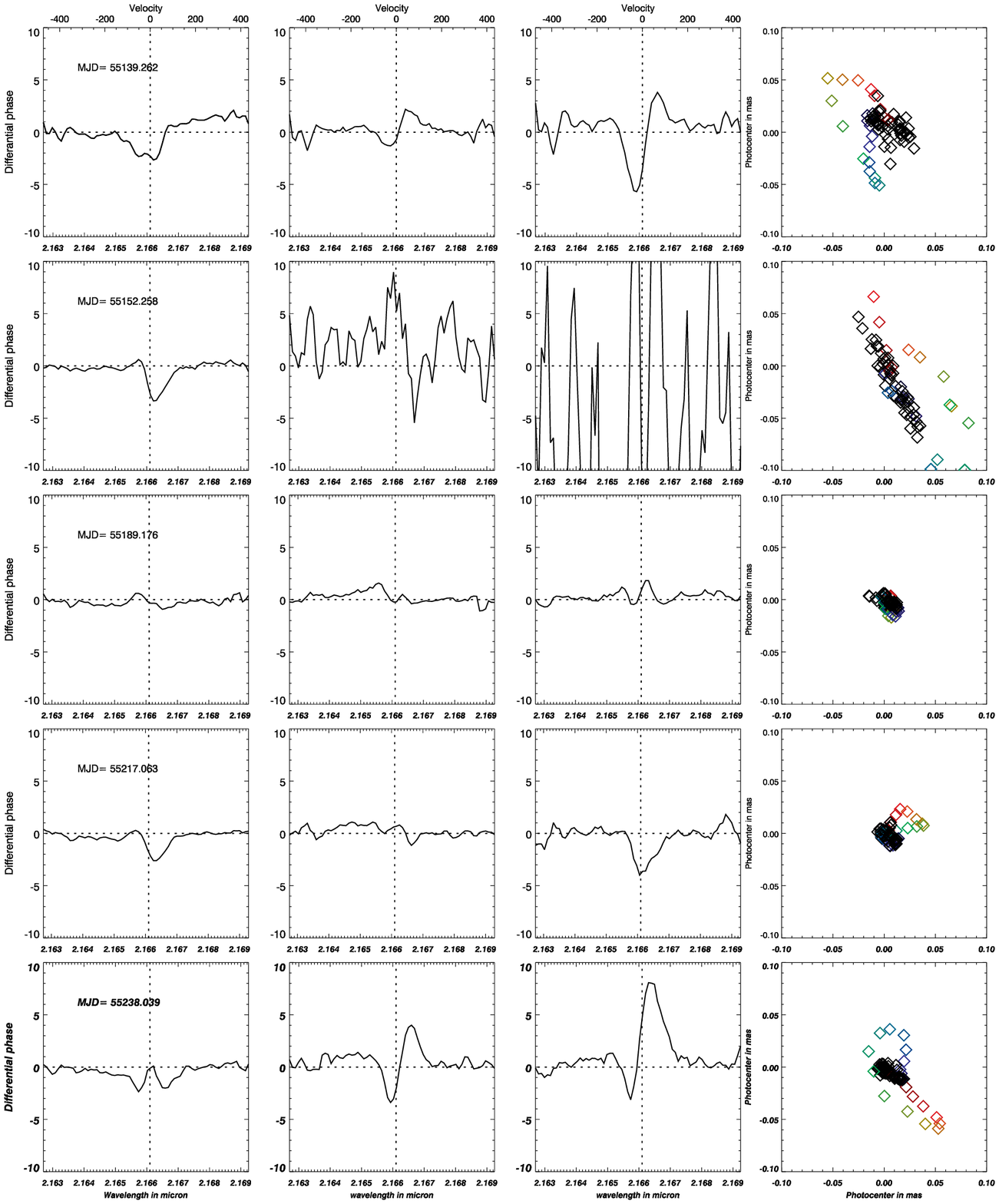}
   \caption{Differential phases and photocenter shift of \object{Rigel} obtained in 2009-2010.}
   \label{fig:Rigel_phase2009} 
  \end{figure*}
  \begin{figure*}[h!]
   \centering
      \includegraphics[width=0.98\textwidth]{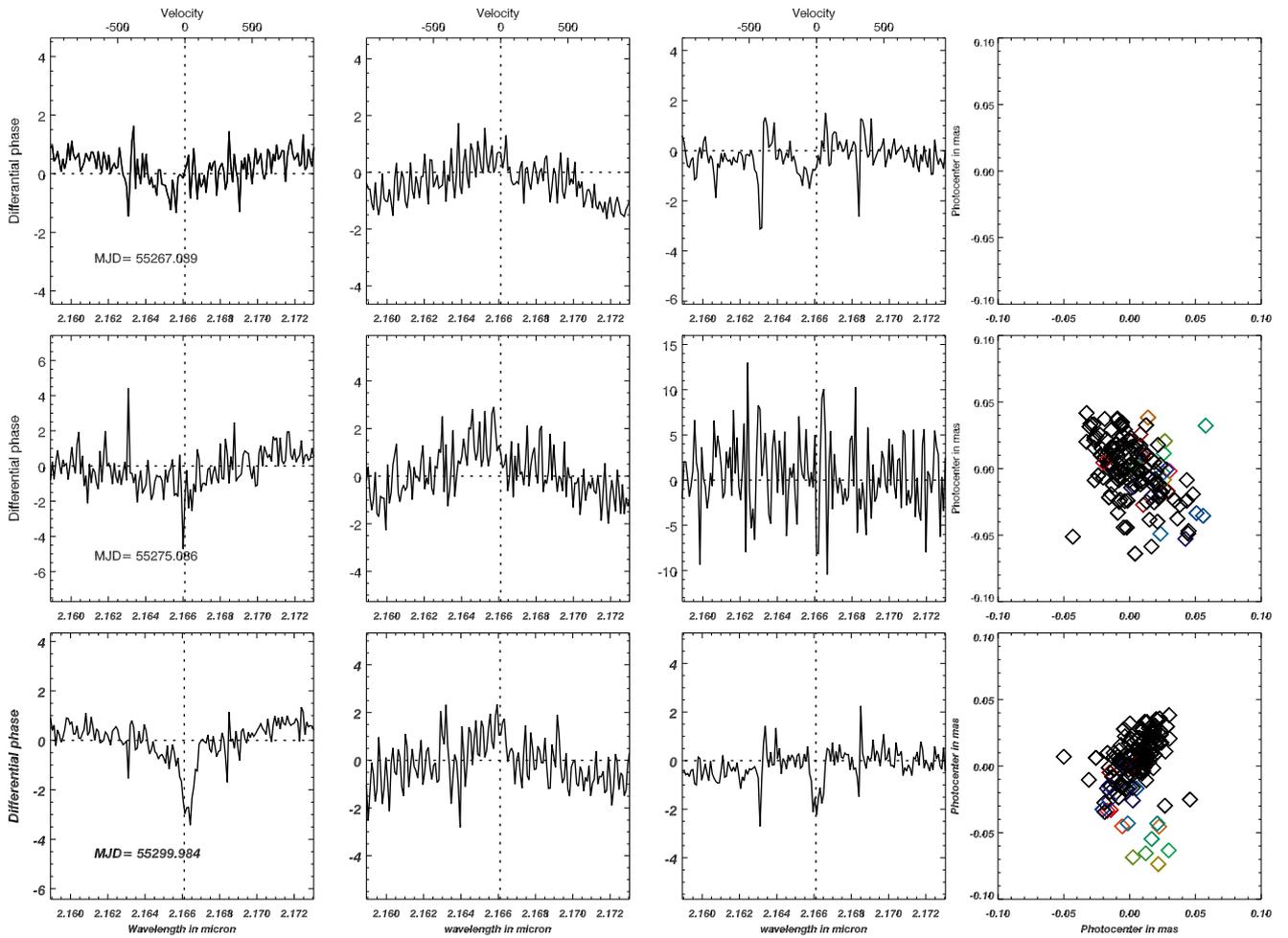}
   \caption{Differential phases and photocenter shift of \object{Rigel} obtained in 2009-2010 (continued). At MJD = 55267.099 (upper row), the baselines were aligned preventing the computation of a 2D photocenter. }
   \label{fig:Rigel_phase2010} 
  \end{figure*}
\end{document}